\RequirePackage[2020-02-02]{latexrelease}
\documentclass[12pt,preprint]{revtex4}
\usepackage{graphicx}
\usepackage{amsmath}
\usepackage{amsfonts,amssymb}
\usepackage[matrix,arrow,curve,frame,poly,arc]{xy}
\usepackage[english]{babel}

\def\be{\begin{equation}}
\def\ee{\end{equation}}
\def\ba{\begin{eqnarray}}
\def\ea{\end{eqnarray}}

\begin{document}
\bibliographystyle{plainnat}

\title{Exact traveling wave solutions of one-dimensional models of cancer invasion}

\author{Maria Shubina }

\email{yurova-m@rambler.ru}

\affiliation{Skobeltsyn Institute of Nuclear Physics\\Lomonosov Moscow State University
\\ Leninskie gory, GSP-1, Moscow 119991, Russian Federation}


\begin{abstract}

In this paper we consider continuous mathematical models of tumour growth and invasion based on the model introduced by Chaplain and Lolas for the case of one space dimension. The models consist of a system of three coupled nonlinear reaction-diffusion-taxis partial differential equations describing the interactions between cancer cells, the matrix degrading enzyme and the tissue. For these models under certain conditions on the model parameters we obtain exact analytical solutions in terms of traveling wave variables. These solutions are smooth positive definite functions for some of which whose profiles agree with those obtained from numerical computations \cite{Chaplain&Lolas2006} for not very large time intervals. 

\end{abstract}

\keywords{partial differential equation, exact solution, traveling wave solutions, cancer invasion, chemotaxis, haptotaxis}

\maketitle

\section{Introduction}

The creation of mathematical models for the description of dynamic biological processes has a long history. Since its inception, mathematical models in oncology have been actively developing. The phenomenon of the onset and development of cancerous tumours is very complicated, it includes many interrelated processes at many spatial and temporal scales. Describing schematically, a solid tumour grows through two distinct phases: the initial phase is the avascular phase, and then the vascular phase. During the avascular growth phase the size of the solid tumour is restricted and the tumour remains localised. The transition from avascular growth to vascular one depends on the formation of blood vessels from a pre-existing vasculature, or angiogenesis. During the growth of a solid tumour the capillary sprout network is formed in response to chemical stimuli (tumour angiogenic factors) secreted by the cells of this tumour \cite{Folkman&Klagsbrun1987} - \cite{Anderson&Chaplain2000}. This stimulates neighboring capillary blood vessels to grow and penetrate the tumour, repeatedly supplying this tumour with a vital nutrient. With the vascular growth phase the process of cancer invasion of peritumoral tissue can and does take place \cite{Chaplain&Lolas2006}. Invasion of tissue plays a key role in the growth and spread of cancer and the formation of metastases.  The process of invasion is the following: the cancer cells secrete various matrix degrading enzymes, which destroy the surrounding tissue or extracellular matrix (ECM), and then the cancer cells actively spread into the surrounding tissue through migration and proliferation. Thus, the primary tumour locally invades the surrounding tissue and spreads to distant sites of the body to form secondary tumours \cite{Enderling&Chaplain2014}. These are secondary tumours (metastases) which are the main cause of death from cancer. 

The  exhaustive exposition of the biological and medical aspects underlying the construction of mathematical models can be found in the cited literature. The study of the various phases of solid tumour growth by mathematical modeling methods has been going on for several decades. A comprehensive review of this area is given in the works \cite{Adam&Bellomo}-\cite{Lowengrub} and references therein. Mathematical modeling of cancer invasion is also presented in the research  literature. One of the first models of cancer cells invasion, described by the system of partial differential equations, were developed in \cite{Gatenby&Gawlinski} and in \cite{Perumpanani}, \cite{Anderson&Chaplain2000}; in the last two works the continuous mathematical models are described by the systems of reaction-diffusion-taxis partial differential equations, where the processes of haptotaxis and chemotaxis, or directed cells movement in response to chemical concentration gradient (gradient of extracellular matrix density or of adhesive molecules in the extracellular matrix) plays a key role in the tumour cell migration. 
As known the macroscopic classical model of chemotaxis was proposed by Patlak in 1953 \cite{P} and by Keller and Segel in 1970s \cite{KS1}-\cite{KS3}, and at present the key role of chemotaxis and haptotaxis in cancer invasion models is beyond doubt and continues to be actively studied. So, in the review \cite{P2018} Painter writes that "Chemotaxis is a fundamental guidance mechanism of cells and organisms ... the Patlak-Keller-Segel (PKS) system forms part of the bedrock of mathematical biology" and "Applications of PKS systems are considered in their diverse areas, including microbiology, development, immunology, cancer ...".

The model introduced in Anderson \textit{et al} \cite{Anderson&Chaplain2000} consists of three partial differential equations and describes the space-time evolution and interaction of cancer cells, the matrix degrading enzyme (the urokinase-type plasminogen activator, uPA) and the host tissue. Further modifications and development of this model was obtained by Anderson \cite{Anderson2005}, Chaplain and Lolas \cite{Chaplain&Lolas2005}, Enderling \textit{et al} \cite{Enderling2006}; the authors of paper  \cite{Chaplain&Lolas2006} and Andasari \textit{et al} \cite{Andasari&Gerisch&Lolas2010} focus their attention on the role of a generic matrix degrading enzyme (MDEs) such as uPA and metalloproteinases. Gerisch and Chaplain \cite{Gerisch&Chaplain2008} formulate a continuum model incorporating the cell-cell and cell-matrix adhesion using non-local terms; different aspects of cell invasion are also considered in \cite{Frieboes}-\cite{Painter2010}; Enderling and Chaplain \cite{Enderling&Chaplain2014} describe fundamentals of mathematical modeling of tumour growth and summarize most prominent approaches. Currently, the invasion models continue to actively develop: in \cite{Peng&Trucu} a new two-scale moving boundary model of cancer invasion is presented; in \cite{Domschke&Trucu} the authors establish a general spatio-temporal-structural framework that allows to describe the interaction of cell population dynamics with molecular binding processes;  a nonlocal mathematical model describing cancer cell invasion as a result of integrin-controlled cell-cell adhesion and cell-matrix adhesion is developed in \cite{Bitsouni&Chaplain}, \cite{Bitsouni&Trucu}; in \cite{Szymanska&Cytowski} the authors present two mathematical models related to different aspects and scales of cancer growth. In \cite{PW2017} - \cite{KZ2018} the authors study the coupled parabolic chemotaxis–haptotaxis system with remodeling of non-diffusible attractant and prove the global existence and uniqueness of classical solutions for a positive logistic growth rate of cancer cells. In the paper \cite{BPFVAPS2019} Bubba \textit{et al} present new results of three-dimensional in vitro cultures of breast cancer cells exhibiting patterns and propose that the main mechanism which leads to the emergence of patterns is chemotaxis. Studying a Keller–Segel PDE system to model chemotactical auto-organization of cells, the authors prove that it admits Turing unstable solutions under a time-dependent condition. A coupled chemotaxis–haptotaxis model of cancer invasion with or without kinetic source in a $2D$ bounded and smooth domain is considered in \cite{XZ2019}. For a large class of cell kinetic sources the authors detect explicit conditions to ensure uniform-in-time boundedness for the Neumann problem. Tao and Winkler \cite{TW2020} study a haptotaxis system proposed as a model for oncolytic virotherapy that considers the interaction between uninfected cancer cells, infected cancer cells, extracellular matrix (ECM) and oncolytic virus. The authors establish a global classical solvability of  the initial-boundary value problem in one- or two-dimensions with a convenient choices of initial data.

The existence of traveling wave solutions for different models of tumour invasion with haptotaxis term was established in the numerical computations results \cite{Chaplain&Lolas2006}, as well as in  \cite{Perumpanani&Sherratt1999}-\cite{Harley} where a detailed study of travelling wave behaviour is performed. However, as far as we know, the models considered in this paper and the solutions presented here are new.

As mentioned above, the model considered here is based on the continuous mathematical model of generic solid tumour growth and invasion introduced in \cite{Chaplain&Lolas2006}. In this work the authors initially develop and modify the mathematical model of Anderson, Chaplain \textit{et al} suggested in 2000 \cite{Anderson&Chaplain2000}. Chaplain and Lolas first focused solely on the interactions between the cancer cells and the surrounding tissue, and study a system of three coupled nonlinear partial differential equations (PDEs) describing the space-time behavior of tumour cells, ECM density and uPA protease concentration. 

First it is assumed that the cell number density $ c(t,\overrightarrow{r}) $ changes because of dispersion arising from the random locomotion. The choice of the cell random motility coefficient $D_{c}$ to be constant does not affect the general framework of the process since the contribution of the chemokinetic term is always the smallest in the locomotion of cancer cells \cite{Chaplain&Lolas2006}. Further the cell density is assumed to be also changed due to directed migratory response of tumour cells to gradients of diffusible (uPA) and non-diffusible (ECM) macromolecules, or chemotaxis and haptotaxis respectively, see \cite{Chaplain&Lolas2006} and references therein. Regarding the proliferation of cancer cells, it is assumed that in the absence of any extracellular matrix, cancer cell proliferation satisfies a logistic growth law; the presence of ECM leads to competition for space between the cancer cells and the ECM \cite{Chaplain&Lolas2006}. This distinguishes the model of Chaplain and Lolas from the one proposed by Anderson, Chaplain \textit{et al} \cite{Anderson&Chaplain2000} since they do not consider any cell proliferation in order for to focus completely on cell-matrix interactions.

The ECM $ v(t,\overrightarrow{r}) $  is not assumed to move and it changes solely due to its degradation by uPA protease upon contact and to its remodelling by cancer and other cells. 

The behavior and the evolution of uPA protease concentration $ u(t,\overrightarrow{r}) $ is governed by diffusion, protease production and protease decay. uPA is produced by the tumour cells, diffuses throughout the extracellular matrix with constant diffusion coefficient $ D_{u} $ and undergoes decay.

In order to solve the system numerically, Chaplain and Lolas non-dimensionalise the equations. Thus the complete system of dimensionless equations describing the interactions between the tumour cells $ c $, extracellular matrix $ v $ and uPA $ u $ has the form \cite{Chaplain&Lolas2006}:
\be 
\left\{
\begin{aligned}
c_{t} & =  D_{c}\, \nabla^{2} c - \chi_{c} \, \nabla (c \nabla u) - \xi_{c} \, \nabla(c \nabla v) + \mu_{1} \, c \, (1 - c - v ) \\ 
v_{t} & =   - \delta u \, v + \mu_{2} \, v \,(1 - c - v )  \\
u_{t} & =  D_{u}\, \nabla^{2} u + \alpha c  - \beta u \\
\end{aligned}
\right.
\ee
where the constant positive parameters of the model are: $ D_{c} $ and $ D_{u}  $ are cells and uPA diffusion coefficients respectively, $ \chi_{c}$ and $ \xi_{c}$ are chemotaxis and haptotaxis coefficients; $ \delta $ is the rate of ECM degradation by uPA; $ \alpha $ and $ \beta $ are the rates of uPA production and decay. The transformation of the variables and parameters in equations (1) into dimensionless quantities is the same as in the cited works:
\ba
\nonumber
t & \rightarrow & \frac{t}{\tau}, \,\, x \rightarrow \frac{x} {L}, \,\, c \rightarrow \frac{c}{c_{0}}, \,\, v \rightarrow \frac{v} {v_{0}}, \,\, u \rightarrow \frac{u} {u_{0}}; \\ 
\nonumber
D_{c} & \rightarrow & \frac{D_{c}}{ D}, \,\, D_{u} \rightarrow \frac{D_{u}} {D},
 \,\, \chi \rightarrow \chi_{c} \frac{u_{0}}{ D}, \,\, \xi \rightarrow \xi_{c}  \frac{v_{0}}{ D}, \,\, \alpha \rightarrow \alpha \tau \frac{c_{0}}{u_{0}}, \,\, \beta \rightarrow \beta \tau, \,\, \delta \rightarrow \delta \tau u_{0}, \\ 
\nonumber 
\mu_{1} & \rightarrow &  \mu_{1} \tau, \,\, \mu_{2} \rightarrow  \mu_{2} \tau, 
\ea
where $ c_{0} $, $ v_{0} $ and $ u_{0} $ are appropriate reference tumour cell density, extracellular matrix density and reference uPA concentration respectively; $ \tau = \frac{L^{2}}{D} $, $ L = 0.1-1 cm $ is the maximum invasion distance of the cancer cells at the early stage of invasion and $ D = 10^{-6} cm ^{2 } s ^{-1}$ is a reference chemical diffusion coefficient 
\cite{Chaplain&Lolas2006}. Since it is assumed that the tumour cells and the matrix degrading enzymes remain within the domain of tissue under consideration \cite{Anderson&Chaplain2000} the new variable $ x \in [0; 1]$. The new parameters are denoted by the same symbols. Consequently the values of parameters are: $ D_{c} \sim 10^{-5} - 10^{-3} $, $ D_{u} \sim 10^{-3} - 1 $, $ \chi_{c} \sim \xi_{c} \sim 10^{-3} - 1$, $ \alpha \sim 0.05 - 1 $, $ \beta \sim 0.13 - 0.95 $, $ \delta \sim 1 - 20 $, $ \mu_{1} \sim 0.05 - 2 $ and $ \mu_{2} \sim 0.15 - 2.5 $.

The model under consideration and its generalizations are actively and fruitfully studied. The results obtained from numerical computations carried out on this model equations produce dynamic, heterogeneous spatio-temporal solutions and demonstrate the ability of the above model to produce complicated dynamics associated with tumour heterogeneity and cancer cell progression and invasion \cite{Chaplain&Lolas2006}. 

It became interesting whether it is possible to solve this system analytically and to obtain exact solutions that allow a satisfactory interpretation. However, it seems that, unfortunately, model (1) cannot be solved analytically even in the case of one space dimension. The aim of this paper is to consider models that, as it seems to us, are as close as possible to the model in \cite{Chaplain&Lolas2006} and which can be solved exactly. We will consider the above systems in one space dimension. In order that the presented results can be compared with the numerical results of the authors cited we study the systems under the assumption that they are  dimensionless. For certain values of the model parameters we obtain exact analytical solutions in terms of traveling wave variable for the velocity depending on these parameters. In cases where these solutions are biologically acceptable we present these solutions and verify that they are consistent with the numerical solutions obtained in \cite{Chaplain&Lolas2006}. At the very end, we will also briefly show other solutions obtained formally and not suitable for biological analysis; perhaps they may be of interest simply as the exact solutions of systems of partial differential equations.

\section{Models under consideration and exact solutions}
\label{sec:1}

\subsubsection{Model without proliferation}

We would like to begin with a model based on the  \cite{Anderson&Chaplain2000} model. Thus, we examine this model with logarithmic chemotactical and haptotactical sensitivity functions, without proliferation and re-establishment terms and with slightly modified equation for ECM. It can be seen that the mathematical model of cancer (1) is more complicated and contains the proliferation term. Nevertheless at first it seems interesting to exactly solve the system where "cell proliferation was not included in order to focus solely on the role of cancer cell migration in invasion" \cite{Enderling&Chaplain2014}. As will be seen from the following the profiles of travelling wave solutions are similar to those in \cite{Chaplain&Lolas2006}, Fig.10 for not very large time intervals.

So, the general form of the considered model is:
\be 
\label{eq:2} 
\left\{
\begin{aligned}
c_{t} & =  D_{c}\, c_{xx} - \chi_{c} \, (c \frac{u_{x}}{u})_{x} - \xi_{c} \, (c \frac{v_{x}}{v})_{x}  \\ 
v_{t} & =   - \delta u \, v^{p}  \\
u_{t} & =  D_{u}\, u_{xx} + \alpha c  - \beta u \\
\end{aligned}
\right.
\ee
where the variables and model parameters are defined above. The second equation in (2) differs from \cite{Anderson&Chaplain2000},  \cite{Chaplain&Lolas2006} by the presence of the power $ p $ of $ v $ and we consider $ 0 < p < 1 $. For $ p = 1 $  we did not obtain biologically acceptable solutions, however, as we shall see below, $  p $ can be made close to $ 1 $, for example $ p = 0.95 $. Here the transformation of the variables and parameters into dimensionless quantities is the same as above except $ \delta $, $ \chi_{c} $ and $ \xi_{c} $: $ \delta \rightarrow \delta \tau u_{0} {v_{0}}^{p - 1}$, so that $ \delta \sim 10^{-5(p-1)} - 20 \times 10^{-6(p-1) }  $; as for $ \chi_{c} $ and $ \xi_{c} $ we take the dimensionless values as in \cite{Chaplain&Lolas2006} i.e $ \chi_{c} \sim \xi_{c} \sim 10^{-3} - 1$.

In terms of traveling wave variable $ y = x - \nu t $, $ \nu = const $ this system has the form:
\be 
\left\{
\begin{aligned}
\nu \, c + D_{c}\, c_{y} - \chi_{c} \, c (\ln u)_{y} - \xi_{c} \, c (\ln v)_{y}  = \lambda \\ 
\nu \, v_{y} - \delta u \, v^{p}   = 0 \\
\nu \, u_{y} + D_{u}\, u_{yy} + \alpha c  - \beta u = 0, \\
\end{aligned}
\right. 
\tag{\ref{eq:2}$*$} 
\ee
where $ c = c(y) $, $ v=v(y) $, $ u = u(y) $ and $ \lambda $ is an integration constant. Further we put $ \lambda = 0 $. If we introduce the function 
\be
F = \dfrac{v ^{1-p}}{1-p},
\ee
the first two equations in ($2*$) give
\be
\left\{
\begin{aligned}
c & = C_{c} \, (e^{-\nu y} \, v^{\xi_{c}} \, u^{\chi_{c}})^{\frac{1}{D_{c}}} \\
u & = \frac{\nu}{\delta} \, F_{y},
\end{aligned}
\right. 
\ee
$ C_{c} > 0 $ is a constant. Substituting (3) and (4) into the third equation of system ($2*$) we obtain:
\be
D_{u} F_{yyy} + \nu F_{yy} - \beta F_{y} + C_{1} \, e^{-\frac{\nu}{D_{c}}y}\, F_{y} ^{\frac{ \chi_{c} }{D_{c}}} \, F^{\frac{\xi_{c} }{D_{c} (1-p)}} = 0,
\ee 
where $ C_{1} = C_{c} \alpha (\frac{\nu}{\delta})^{\frac{\chi_{c}}{D_{c}} -1} (1 - p)^{\frac{\xi_{c} }{D_{c} (1-p)}} $, and further we will investigate this equation. 

It seems to us that for arbitrary values of the system parameters it is impossible to obtain an exact solution in explicit form. Therefore, we impose a number of restrictions on these parameters. So, let 
\be
\frac{\chi_{c} }{D_{c}} = 1
\ee
that is, we are considering $ \chi_{c} = D_{c} \sim 10^{-3} $. Then there is a 'chosen' value of the speed of the traveling waves for which we obtain two classes of different solutions. Let 
\be
\nu ^{2} = \dfrac{\beta  D_{c}^{2} }{D_{u} - D_{c}}
\ee
It is possible to do this because $ D_{u} \geq D_{c} $ and we do not consider the case $ D_{u} = D_{c} $. Then it can be shown that equation (5) can be reduced to the form: 
\ba
F_{yy} - \frac{\nu (D_{u} - D_{c})}{D_{u} D_{c}} \, F_{y} + \frac{C_{1} D_{c} (1-p)}{D_{u} (\xi_{c} + D_{c} (1-p))}  \, e^{-\frac{\nu}{D_{c}}y}\, F ^{\frac{\xi_{c} }{ D_{c} (1-p)} \, + 1} = 0
\ea 
(with a constant of integration equal to zero). To integrate this equation we use the Lie group method of infinitesimal transformations \cite{Olver}. We find a group invariant of a second prolongation of one--parameter symmetry group vector of (8) and with its help we transform equation (8) into an equation of the first order. It turns out that there are two nontrivial symmetry groups depending on the ratio of the parameters that give two different types of solutions. Let us consider the first of them.

\subsubsection{Exact solution}

The possibility to reduce equation (8) to a first order equation and solve it requires the following condition:
\be
1 - p = \dfrac{\xi_{c} (D_{u} - D_{c}) }{2 D_{u} D_{c} }. 
\ee
If we introduce the new variable $ z $ and the new function $ w $:
\ba
z & = & F \,e^{- \frac{\nu (1 - p) }{\xi_{c} } \,\, y}\\
\nonumber
w & = & - \dfrac{D_{c}}{\nu} \, F_{y} \, e^{- \frac{\nu (1 - p)}{\xi_{c} } \,\, y}
\ea
then after elementary integration equation (8) turns into a quadratic equation on $ w(z) $:
\be
w^{2} + \frac{2 (1-p) D_{c}}{\xi_{c}} \, z \, w + C_{2} \, z^{\frac{\xi_{c} }{ D_{c} (1-p)} \, + 2} = 0, 
\ee
where the constant of integration is equal to zero and $ C_{2} = \frac{2 C_{c} D_{c}^{4} (1-p)^{\frac{\xi_{c} }{ D_{c} (1-p)} + 2}}{\nu^{2}  (\xi_{c} + D_{c} (1-p))  (\xi_{c} + 2 D_{c} (1-p))} $. Returning to the initial function $ F $ and variable $ y $ and integrating (11) we obtain a solution for $ F $:
\be
F = C_{3} \,\, (e^{-\frac{\nu}{D_{c}} \,\, y} + C_{F})^{- \frac{2 D_{c} (1-p)}{\xi_{c}}}
\ee
where $ C_{F} $ is a positive constant and $ C_{3} = ( C_{F} \frac{ 2 \nu^{2}  D_{u} (\xi_{c} + D_{c} (1-p))  (\xi_{c} + 2 D_{c} (1-p)) (1-p)^{-\frac{\xi_{c} }{ D_{c} (1-p)} }}{ C_{c} \alpha D_{c}^{2} \xi_{c}^{2}} )^{\frac{ D_{c} (1-p)}{\xi_{c}}} $. Substituting this into (3) and (4) we obtain the first type of solutions of system (2) in the form:
\ba
\nonumber
c(y) & = & {C_{C}} \, e^{-\frac{2 \nu}{D_{c}} \,\, y} \,\,  (e^{- \frac{\nu}{D_{c}} \,\, y} + C_{F})^{- \frac{D_{u} (\xi_{c} + 2 D_{c} (1-p))}{\xi_{c}}\,\, - 2}
 \\
v(y) & = & {C_{v}} \, (e^{-\frac{\nu}{D_{c}} \,\, y} + C_{F})^{- \frac{2 D_{c} }{\xi_{c}}}
 \\
\nonumber
u(y) & = & {C_{u}}\, e^{-\frac{\nu}{D_{c}} \,\, y} \,\,  (e^{-\frac{\nu}{D_{c}} \,\, y} + C_{F})^{- \frac{D_{u} (\xi_{c} + 2 D_{c} (1-p))}{\xi_{c}}},
\ea
where the constants are
\ba
\nonumber
C_{v} & = & \left(  \frac{C_{F}}{C_{c}} \, \frac{\beta  (\xi_{c} + D_{c} (1-p))  (\xi_{c} + 2 D_{c} (1-p)) }{\alpha D_{c} \xi_{c} (1-p)} \right) ^{\frac{D_{c}}{\xi_{c}}}
\\
C_{u} & = & {C_{v}}^{\frac{1}{1-p}} \frac{\beta D_{c}}{\delta  D_{u}  (1-p)}  \\
\nonumber
{C_{C}} & = & C_{c}\, {C_{u}} \, {{C_{v}}}^{\frac{\xi_{c}}{D_{c}}}. 
\ea 

As can be seen from (13) these solutions are the positive functions defined for all values of $ y $. Despite the fact that because of the biological context we are interested in solutions in a restricted space-time domain, it is easy to see that for $ D_{u} > D_{c} $ the tumour cell density $ c(y) $ and the uPA concentration $ u(y) $ vanish at $ y \rightarrow \pm \infty $; the extracellular matrix density $ v(y) \rightarrow  C_{v} C_{F}^{- \frac{2 D_{c} }{\xi_{c}}} $ at $ \nu y \rightarrow + \infty $ and $ v(y) \rightarrow  0 $ at $ \nu y \rightarrow - \infty $. It can also be seen that $ c(y) $ and $ u(y) $ have a single maximum; its values, as well as the asymptotic value of $ v(y) $ at $ \nu y \rightarrow + \infty $ depend on the chosen parameters. These functions are presented in Fig.1--Fig.2 for different values of parameters and for $\nu > 0$. In Fig.3 we show the sequence of profiles of travelling waves which propagate into the tissue. A thick line shows the curves at the time $ t = 0 $.
\begin{figure}[h!]
\begin{minipage}{0.29\linewidth}
\center{\includegraphics[width=1\linewidth]{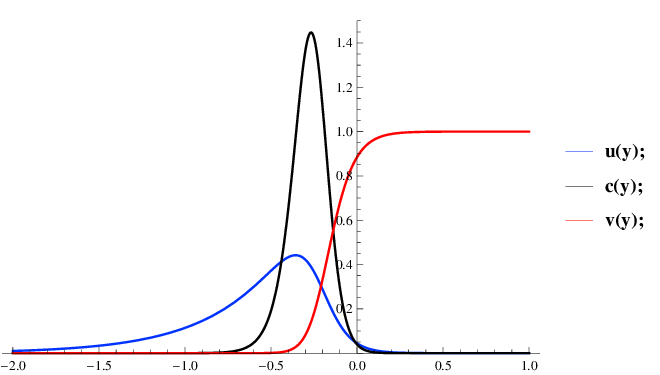} \\ Fig.1: $ p = 0.9 $; $ D_{c} = \chi_{c} = 8\times 10^{-3} $; $D_{u} = 10^{-2} $; $\xi_{c}= 8\times 10^{-3}$; $ \alpha = 0.25 $; $ \beta= 0.3 $; $ \delta = 10^{0.5} $; $ C_{F} = 15.84 $; $ C_{c} = 1$}
\end{minipage}
\hfill
\begin{minipage}{0.29\linewidth}
\center{\includegraphics[width=1\linewidth]{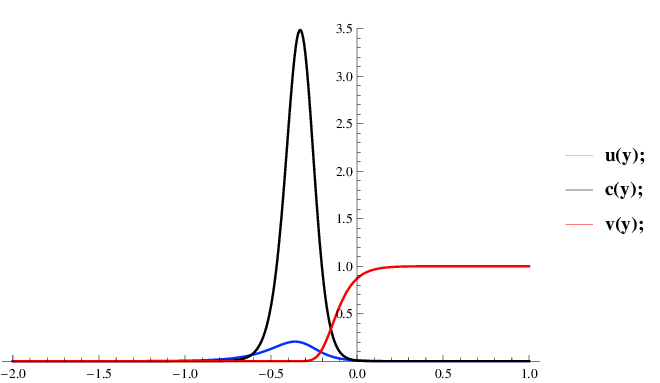} \\ Fig.2:  $ p = 0.95 $; $ D_{c} = \chi_{c} = 5\times 10^{-3} $; $D_{u} = 10^{-2} $; $\xi_{c}= 10^{-3}$; $ \alpha = 0.1 $; $ \beta= 0.95 $; $ \delta = 10^{1.25} $; $ C_{F} = 71.25 $; $ C_{c} = 1$}
\end{minipage}
\begin{minipage}{0.32\linewidth}
\center{\includegraphics[width=1\linewidth]{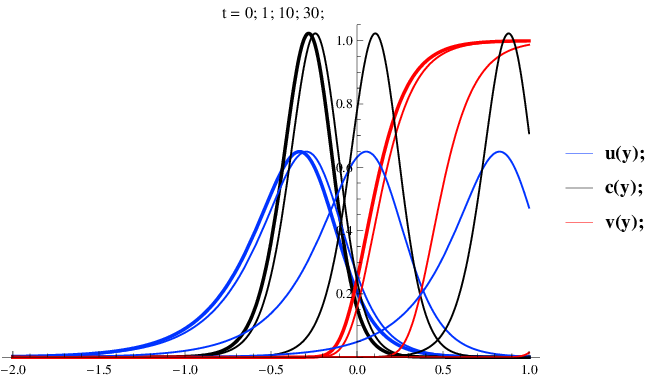} \\ Fig.3:  $ p = 0.95 $; $ D_{c} = \chi_{c} = 5\times 10^{-3} $; $D_{u} = 10^{-2} $; $\xi_{c}= 10^{-3}$; $ \alpha = 0.34 $; $ \beta= 0.3 $; $ \delta = 10^{0.25} $; $ C_{F} = \frac{225}{34} $; $ C_{c} = 1$}
\end{minipage}
\hfill
\end{figure}

\subsubsection{Model with proliferation}

Let us consider now the model with proliferation and re-establishment terms, modified as follows:
\be 
\label{eq:15} 
\left\{
\begin{aligned}
c_{t} & =  D_{c}\, c_{xx} - \chi_{c} \, \big( c ( u_{x} + \lambda_{u} u)\big)_{x} - \xi_{c} \, (c \frac{v_{x}}{v^{p}})_{x} + \mu_{1} \, c \, (1 - c ) \\ 
v_{t} & =   - \delta u \, v^{p} + \mu_{2} \, v^{p} \,(1 - \lambda_{c} \,c - v^{1-p} )  \\
u_{t} & =  D_{u}\, u_{xx} + \alpha c  - \beta u \\
\end{aligned}
\right.
\ee
where we again entered the power $ p $ of $ v $, $ 0 < p < 1 $ and the constant $ \lambda_{c} > 0 $; other variables and model parameters are defined above, except $ \mu_{2} $: $ \mu_{2} \rightarrow \mu_{2} \tau {v_{0}}^{p - 1}$, so that $ \mu_{2} \sim 0.15 \times 10^{-5(p-1)} - 2.5 \times 10^{-6(p-1) }  $. But the main difference between system (1) from this one is the presence of the additional term $ \lambda_{u} ( c u)_{x} $, $ \lambda_{u}$ is a constant. This term was initially added to the first equation (15) in order to be able to solve the system exactly. However the profiles of solutions obtained are very close to the profiles in Figure 4--Figure 5 \cite{Chaplain&Lolas2006} of travelling wave which propagates into the tissue. This suggests that the addition of the above term does not greatly distort the original model despite the fact that the coefficient $ \lambda_{u} $ cannot be made arbitrarily small; as can be seen from the following $ \lambda_{u} \geq D_{c} $.

In terms of traveling wave variable $ y = x - \nu t $, $ \nu = const $ this system has the form:
\be 
\left\{
\begin{aligned}
D_{c} \, c_{yy} + \nu c_{y} - \chi_{c} \, \big( c ( u_{y} + \lambda_{u} u)\big)_{y} - \xi_{c} \, (c \frac{v_{y}}{v^{p}})_{y} + \mu_{1} \, c \, (1 - c ) = 0 \\ 
\nu  v_{y} - \delta u \, v^{p} + \mu_{2} \, v^{p} \,(1 - \lambda_{c} \,c - v^{1-p} ) = 0 \\
D_{u}\, u_{yy} + \nu u_{y} + \alpha c  - \beta u = 0\\
\end{aligned}
\right.
\tag{\ref{eq:15}$*$} 
\ee
where $ c = c(y) $, $ v=v(y) $, $ u = u(y) $. Everywhere below we consider $ \nu > 0 $.

As in the case of the previous model ($2*$) we must impose a number of conditions on model parameters to obtain exact solutions in explicit form. So, the power $ p $ and the constant $ \lambda_{u} $ are expressed through  other model parameters by the relations:
\be
1 - p  =  \frac{\xi_{c} \,(\delta \, \alpha + \beta \, \mu_{2} \, \lambda_{c} )}{\mu_{2} \,(\chi_{c} \, \alpha + \xi_{c} \,\mu_{2} \, \lambda_{c})}, \,\,\,\,
\lambda_{u}  =  \frac{\beta \,\sqrt{\xi_{c} \, \mu_{2} \, \lambda_{c} \,- \,D_{c} \, \mu_{1} } }{\chi_{c} \, \alpha},
\ee
where we are restricted by the condition $ \xi_{c} \, \mu_{2} \, \lambda_{c} \,- \,D_{c} \, \mu_{1} > 0 $. The 'chosen' value of the speed of the traveling waves is
\be
\nu = \frac{\xi_{c} \, \mu_{2} \, \lambda_{c}}{\sqrt{\xi_{c} \, \mu_{2} \, \lambda_{c} \,- \,D_{c} \, \mu_{1} }}.
\ee
And one more necessary condition that must be imposed on the model constants in order to obtain exact solutions has the form:
\be
D_{u} \, (\delta \, \alpha + \beta \, \mu_{2} \, \lambda_{c} ) \, (\xi_{c} \, \mu_{2} \, \lambda_{c} \,- \,D_{c} \, \mu_{1} )\, = \, \chi_{c} \, \alpha \,\mu_{2} \, \lambda_{c} \,(\chi_{c} \, \alpha + \xi_{c} \,\mu_{2} \, \lambda_{c}).
\ee
Then the second equation of ($15*$) can be integrated and we obtain
\be
v(y) = \Big(  1 -  \frac{\mu_{2} \, \lambda_{c} \,D_{u} \,(1 - p) }{\alpha \,\nu } \,\, u_{y}  - \frac{\delta \, \alpha + \beta \, \mu_{2} \, \lambda_{c}}{\alpha \,\mu_{2}} \,\, u \Big )^{\frac{1}{1-p}}         
\ee
Further we introduce the function 
\be
F = c_{y} + K \, (c - c^{2}),\,\,\,\,K = \frac{\xi_{c} \, \mu_{2} \, \lambda_{c}}{D_{c} \, \nu}.
\ee
Then it can be verified by direct substitution, that first equation ($15*$) can be reduced to the form:  
\ba
F_{y} + \frac{\mu_{1} \, \nu}{\xi_{c} \, \mu_{2} \, \lambda_{c}} \, F = 0.
\ea

\subsubsection{Exact solution}

This first-order linear equation (21) has the obvious solution:
\be
F = C_{F} \, e^{- \frac{\mu_{1} \, \nu}{\xi_{c} \, \mu_{2} \, \lambda_{c}}\,\,y }
\ee
Let us consider the case $ C_{F} = 0 $, or $ F = 0 $. Then the function $ c(y) $ must be a solution of the Riccati equation 
\be
c_{y} + K \, (c - c^{2}) = 0
\ee
where the left side of (23) is in the same time the integrated Burgers equation in a traveling wave variable. Herewith tumour cell density $ c $ is a well known traveling wave solution of the Burgers equation and we are interested in a bounded 'shock wave' one:
\be
c = \frac{1}{1 + C_{c}\, e^{K \,y}},
\ee
where $ C_{c} $ is a positive constant. Substituting this expression in the third equation ($15*$) for uPA concentration and choosing the integration constant so that u(y) is bounded we get: 
\ba
u(y) & = & \frac{\alpha}{D_{u}\,\Delta \textit{k} \, \textit{k}_{+}} \bigg(  C_{c}^{\frac{\textit{k}_{+}}{K}} \, \Gamma (1 - \frac{\textit{k}_{+}}{K}) \, \Gamma ( 1 + \frac{\textit{k}_{+}}{K}) \, e^{\textit{k}_{+}\,y}  - \, _{2}F_{1} ( -\frac{\textit{k}_{+}}{K}, 1, 1 - \frac{\textit{k}_{+}}{K}, -C_{c} \, e^{K\,y} ) \bigg) \\
\nonumber
& + & \frac{\alpha}{D_{u}\,\Delta \textit{k} \, \textit{k}_{-}}  \bigg(  \, _{2}F_{1} ( -\frac{\textit{k}_{-}}{K}, 1, 1 - \frac{\textit{k}_{-}}{K}, -C_{c} \, e^{K\,y} )
 \bigg)
\ea
where
\be
\textit{k}_{\pm} = \frac{1}{2 D_{u}} \big( -\nu \pm \sqrt{\nu^{2} + 4\beta D_{u}} \big), \,\, \Delta \textit{k} = \textit{k}_{-} - \textit{k}_{+};
\ee
$ \Gamma $ and $ _{2}F_{1} $ are the Gamma-function and the hypergeometric Gauss function respectively. Finally substituting (24) in (19) we obtain the explicit expression for the extracellular matrix density $ v(y) $.

It is easy to see from (24), that $ c(y) \rightarrow 0 $ at $ y \rightarrow \infty $ and $ c(y) \rightarrow 1 $ at $ y \rightarrow -\infty $. Also, with a certain choice of model constants the uPA concentration (25) is smooth positive definite functions and $ u(y) \rightarrow 0 $ at $ y \rightarrow \infty $ and $ u(y) \rightarrow \frac{\alpha}{\beta} $ at $ y \rightarrow -\infty $. As for the extracellular matrix density  $ v(y) $ we can see from (19) that $ v(y) \rightarrow 1 $ at $ y \rightarrow \infty $ and $ v(y) \rightarrow 1 - \frac{\delta \, \alpha + \beta \, \mu_{2} \, \lambda_{c}}{\beta \,\mu_{2}}$ at $ y \rightarrow -\infty $ and the choice of model parameters should ensure the condition $ v(y) \geq 0 $. These functions are presented in Fig.4--Fig.5. 
\begin{figure}[h!]
\begin{minipage}{0.33\linewidth}
\center{\includegraphics[width=1\linewidth]{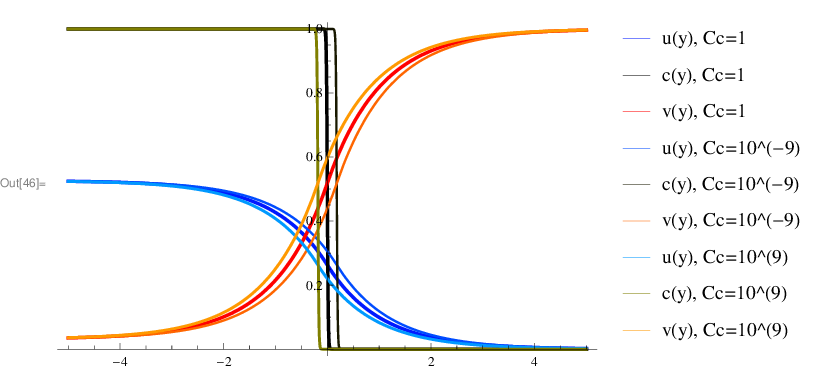} \\ Fig.4: $ p = 0.868 $; $ D_{c} = 10^{-5} $; $D_{u} = 1 $; $ \chi_{c} = 0.475 $; $\xi_{c}= 3.4\times 10^{-2}$; $ \alpha = 0.5 $; $ \beta= 0.95 $; $ \delta = 4.6 $; $ $; $ \mu_{1} = 0.05 $; $ \mu_{2} = 2.5 $}
\end{minipage}
\hfill
\begin{minipage}{0.33\linewidth}
\center{\includegraphics[width=0.9\linewidth]{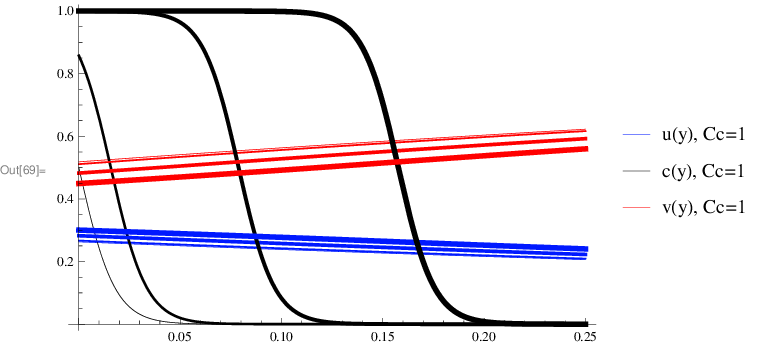} \\ Fig.5: \textit{Sequence of profiles showing the propagation of $ c(y) $, $ u(y) $ and $ v(y) $ into the tissue for propagating $ t=0; 10; 50; 100 $. The parameters are the same as in Fig.4.}}
\end{minipage}
\begin{minipage}{0.32\linewidth}
\center{\includegraphics[width=0.7\linewidth]{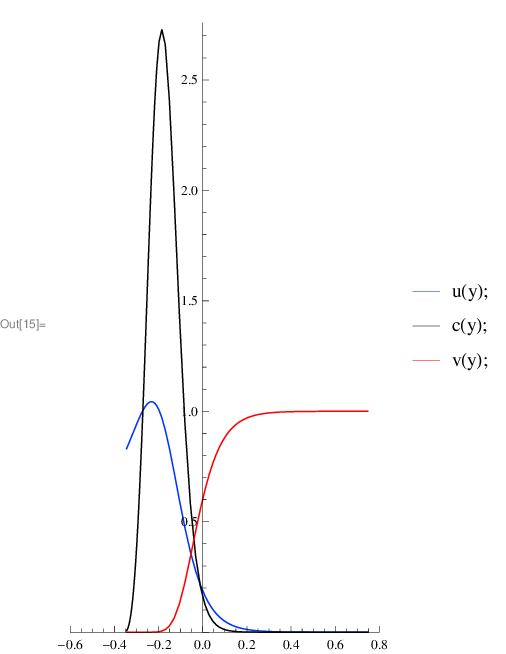} \\ Fig.6:  $ p = 0.9 $; $ D_{c} = \chi_{c} = 8\times 10^{-3} $; $D_{u} = 10^{-2} $; $\xi_{c}= 2.4 \times 10^{-3}$; $ \alpha = 0.5 $; $ \beta= 0.4 $; $ \delta = 4 $; $ C_{w} = 100 $; $ C_{c} = 1$}
\end{minipage}
\hfill
\end{figure}
As is already mentioned, these profiles for the tumour cell density, the protease concentration and the ECM density are very close to the profiles in Figure 4--Figure 5 \cite{Chaplain&Lolas2006} showing the evolution of $ c(x, t) $, $ u(x, t) $ and $  v(x, t) $ "for which successful invasion occurs and a travelling wave which propagates into the tissue is established" \cite{Chaplain&Lolas2006}. In Fig.5 we show the sequence of profiles of propagating travelling waves at different points in time; the parameters  are the same as in Fig.4. The more time moments are considered, the thicker the lines of profiles.

\subsubsection{Unsuitable solutions}

It seems to us that there are no other exact solutions that are acceptable from a biological point of view. Let us now briefly give solutions that are not important for biology but they can be interesting as exact solutions of partial differential equations.

\subsubsection*{Solutions of Model without proliferation}

Let us return to equation (8) and consider the relation similar to (9):
\be
1 - p = \dfrac{\xi_{c} (D_{u} - D_{c}) }{ D_{c} (2 D_{c} - D_{u}) }. 
\ee
It should be noted that (9) and (27) can not be equal because of $ D_{u} > D_{c} $. Further, as can be seen from (14) the restriction on $ D_{u} $ and $ D_{c} $ becomes more strict: $ D_{c}  < D_{u} < 2D_{c} $. Similarly to the case of the first class of solutions we introduce the new variable $ z $ and the new function $ w $:
\ba
z  =  F \,e^{- \frac{\nu (D_{u} - D_{c}) }{D_{u} D_{c} } \,\, y},\,\,\,\,
w  =  F - \dfrac{D_{u} D_{c}}{\nu (D_{u} - D_{c})} \, F_{y} 
\ea
and equation (8) turns into a quadratic equation on $ w(z) $:
\be
w^{2} + C_{4} \, z^{\frac{D_{u} }{ D_{u} - D_{c}} } - C_{w} = 0, 
\ee
where $ C_{w} > 0 $ is the constant of integration, $ C_{4} = \frac{2 C_{c} \alpha (D_{u} - D_{c}) (1-p)^{\frac{\xi_{c}}{D_{c} (1-p)}}}{\beta D_{c}} $. 
Then we find solutions of equation (29) in parametric form whit a parameter $ \tau $:
\be
\tau^{2} + 1 = \frac{C_{w}}{C_{4}}\,\, z^{-\frac{D_{u} }{ D_{u} - D_{c}} }.
\ee
The analysis of solutions asymptotic forms at $ \pm \infty $ \cite{Bateman&Erdelyi} and the requirement of positivity of functions $ c(y) $, $ v(y) $ and $ u(y) $ determine one of the constants of integration. The formulas obtained are rather complicated, for this reason we introduce the notation:
\be
\Theta (\tau) = - \tau \,\, _{2}F_{1} \Big ( \frac{1}{2}; \frac{3}{2} - \frac{D_{c}}{D_{u}}; \frac{3}{2}; - \tau^{2} \Big ) + \frac{\sqrt{\pi} \Gamma(1 - \frac{D_{c}}{D_{u}})}{2 \Gamma (\frac{3}{2} - \frac{D_{c}}{D_{u}})}
\ee
and express our solutions in terms of $ \Theta (\tau) $. We also give the solutions for $ \nu > 0 $. This yields the following expressions for the second type of solutions:
\ba
y(\tau) & = & -\frac{D_{u}}{\sqrt{\beta (D_{u} - D_{c})}}\,\, \ln \Bigg(
\frac{{C_{w}}^{ \frac{1}{2} - \frac{D_{c}}{D_{u} }}}{{C_{4}}^{ 1 - \frac{D_{c}}{D_{u}}}} \,\frac{2 (D_{u} - D_{c})}{D_{u}} \,\,\, \Theta (\tau) \Bigg) 
\ea
\ba
v(\tau) & = & \Big ( \frac{{C_{w}}^{\frac{1}{2}} D_{u} (1 - p)}{2 (D_{u} - D_{c})} \Big )^{\frac{1}{1 - p}} \, (\tau^{2} + 1)^{- \frac{D_{u} - D_{c}}{D_{u} (1 - p)}} \,\,\, \big ( {\Theta (\tau)} \big ) ^{-\frac{1}{1 - p}} 
\ea
\ba
u(\tau) & = & - \frac{{C_{w}}^{\frac{1}{2}} \, \beta D_{c}}{\delta D_{u}}\,\, \tau \, (\tau^{2} + 1)^{-2 + \frac{D_{c}}{D_{u}}} \, \Bigg( \, _{2}F_{1} ( \frac{1}{2}; \frac{3}{2} - \frac{D_{c}}{D_{u}}; \frac{3}{2}; - \tau^{2} ) - \\ 
& - & \tau^{2}  \big( 1 - \frac{2D_{c}}{3D_{u}} \big )\,\, _{2}F_{1} ( \frac{3}{2}; \frac{5}{2} - \frac{D_{c}}{D_{u}}; \frac{5}{2}; - \tau^{2} ) \Bigg)^{-1} \, + \frac{{C_{w}}^{\frac{1}{2}} \beta D_{c}}{2 \delta (D_{u} - D_{c})} \,\, (\tau^{2} + 1)^{-1 + \frac{D_{c}}{D_{u}}} \,\,\,\big ( {\Theta (\tau)}\big ) ^{-1} \nonumber
\ea
and since (4) the expression for the tumour cells density $ c(\tau) $ has the form:
\ba
c(\tau) & = & \frac{2 \beta D_{c} (D_{u} - D_{c})}{\alpha {D_{u}}^{2}} \,\, (\tau^{2} + 1)^{1 - \frac{2 D_{c}}{D_{u}}} \,\, \big ( {\Theta (\tau)}\big ) ^{2} \, u(\tau).
\ea
Here $ _{2}F_{1} $ is the hypergeometric Gauss function, $ \Gamma $ is the Gamma-function. We can see from (31), (33)-(35) that the functions $ v(\tau) $, $ u(\tau) $ and $ c(\tau) $ are smooth positive definite functions for all $ \tau $. The tumour cell density $ c(\tau)  \rightarrow 0 $ at $ \tau \rightarrow \pm \infty $, the uPA concentration $ u(\tau) \rightarrow u_{0} $ at $ \tau \rightarrow - \infty $ and $ u(\tau) $ vanish at $ \tau \rightarrow +\infty $; the extracellular matrix density $ v(\tau) \rightarrow 0 $ at $ \tau \rightarrow - \infty $ and $ v(\tau) \rightarrow v_{0} $ at $ \tau \rightarrow + \infty $ where the values of $ u_{0} $ and $  v_{0} $ can be obtained from (33)-(34). However as can be seen from (31)-(32) $ y \rightarrow y_{0} $ at $ \tau \rightarrow - \infty $, where $ y_{0} $ is a finite value that can be made $ < 0 $ (or $ \geq 0 $) by choosing constants of integration. This leads to the fact that the solutions obtained as functions of $ y $ can be considered only in a limited time interval. In other words since $ x \in [0; 1]$ and $ t \geq 0 $ formally $ y \in [-\nu t; 1 - \nu t]$ (for $ \nu > 0 $). But since $ y $ is bounded from the left by $ y_{0} $, then for each $ y_{0} $ there exists a value of time $ t_{0} $ after which no solutions are defined. The graphs of these solutions are presented in Fig.6.

\subsubsection*{Solutions of Model with proliferation}

Consider again equation (22) and let $ C_{F} \neq 0 $:
\be
c_{y} + K \, (c - c^{2}) = C_{F} \, e^{- M \,y }, \,\,\, M = \frac{\mu_{1} \, \nu}{\xi_{c} \, \mu_{2} \, \lambda_{c}}.
\ee
As expected, the Cole–-Hopf transformation linearizes this equation; introducing the new function $ \tilde{c} $ and the new variable $ \xi $ as
\ba
c & = & -  \frac{1}{K} \, \frac{\tilde{c}_{y}}{\tilde{c}} \\
\xi & = & \frac{2\sqrt{C_{F} K}}{M}  \, e^{- \frac{M}{2}\,y}
\ea
we obtain the Bessel equation for $ C_{F} > 0 $ and the modified Bessel equation for $ C_{F} < 0 $ with the order $ \tilde{\nu}^{2} = \big(\frac{K}{M}\big)^{2} $. We are only interested in a smooth solution and consider the Infeld $ I_{\tilde{\nu}} $ and the Macdonald $ K_{\tilde{\nu}} $ functions. As a result the solution for tumour cell density $ c(y) $ has the form:
\be
c(y) = \frac{1}{2} \, \big( 1 + \frac{|\tilde{\nu}|}{\tilde{\nu}}  \big) + \sqrt{\frac{|C_{F}|}{K}} \, e^{- \frac{M}{2}\,y} \, \frac{C_{I} I_{\tilde{\nu} + 1}(\xi) -  C_{K} K_{\tilde{\nu} + 1}(\xi)}{C_{I} I_{\tilde{\nu}}(\xi) +  C_{K} K_{\tilde{\nu}}(\xi)}.
\ee
It is easy to see that $ c \rightarrow \infty $ at $ y \rightarrow - \infty $. It cannot be stated that for arbitrary values of $ \tilde{\nu} $ the function $ c(y) $ will always be non-negative and will not diverge at $ y \rightarrow \infty $, but it seems possible to correct this by suitable choosing of $ C_{I} $ and $ C_{K} $. Nonetheless, the exponential growth at $ y \rightarrow - \infty $ is a common property of the obtained solutions. Perhaps this could be acceptable, as it could be interpreted as the exponential growth of tumour cells. However, from ($15*$) it can be expected that the protease concentration $ u(y) $ will also grow exponentially at $ y \rightarrow - \infty $, which is rather difficult to interpret. Solving the third equation ($15*$) we obtain integrals that can be taken exactly only for half-integer $ \tilde{\nu} $. In particular for $ \tilde{\nu} = \frac{1}{2} $, or $ \xi_{c} \, \mu_{2} \, \lambda_{c} = 1.5 D_{c} \, \mu_{1} $, with $ C_{I} = \pi C_{K} $ we obtain very simple expressions for $ c(y) $ and $ u(y) $: 
\ba
c(y) & = & \sqrt{\frac{|C_{F}|}{K}} \, e^{- K \,y}\\   \nonumber
u(y) & = & C_{+} \, e^{\textit{k}_{+} \,y} + C_{-} \, e^{\textit{k}_{-} \,y} - \sqrt{\frac{|C_{F}|}{K}} \, \frac{\alpha}{D_{u}(\textit{k}_{+} + K)(\textit{k}_{-} + K)}\,e^{- K \,y},
\ea
$ C_{\pm} $ are constants; one can put one of them or both equal to zero. It is not obvious whether the requirement $ u \geqslant 0 $ and the conditions (16)--(18) can be met simultaneously. But this is not the main problem of this solution. The most unrealistic in it is the expression for the extracellular matrix density $ v(y) $; as can be seen from (19) this function either diverges at $ y \rightarrow - \infty $ or is equal to $ 1 $ for all $ y $. All this gives reason to believe that these solutions do not meet the biological requirements.

The next solution is even less realistic. The expression for $ c(y) $ can easily be obtained by substitution $ c = e^{- K \, y} \, \tilde{c} $, $ \xi_{c} \, \mu_{2} \, \lambda_{c} = 1.5 \, D_{c} \,\mu_{1}  $, but at the same time the function $ u(y) $ cannot be fully integrated exactly. So, 
\be
c(y) = e^{-Ky}\,\sqrt{\frac{|C_{F}|}{K}}\,\,\dfrac{1 - \tilde{C_{c}}\, e^{-2 \, \sqrt{\frac{|C_{F}|}{K}}\,\,e^{-Ky}}}{1 + \tilde{C_{c}}\, e^ {-2 \, \sqrt{\frac{|C_{F}|}{K}}\,\,e^{-Ky}}}. 
\ee
As the solution above this function obviously diverges at $ y \rightarrow - \infty $. However, the expression for the protease concentration contains integrals of the form $ \int \frac{\big( ln(t - \tilde{C_{c}}) \big)^{\frac{\textit{k}_{\pm}}{K}}\,dt}{t} $, which can be expressed through the polylogarithm for restricted set of $ \textit{k}_{+} $ and cannot be taken exactly for $ \textit{k}_{-} $. In addition, we suppose that $ u(y) \nrightarrow $ exponentially at $ y \rightarrow - \infty $ leading to a divergence of $ v(y) $ as in the solution above.

\section{Conclusion}
\label{sec:5}

In this article we investigate the continuous mathematical models of tumour growth and invasion based on the model introduced by Chaplain and Lolas  \cite{Chaplain&Lolas2006} and by Anderson, Chaplain \textit{et al} \cite{Anderson&Chaplain2000} in one space dimension. The models consist of systems of three coupled nonlinear reaction-diffusion-taxis partial differential equations with different kinds of chemotactical, haptotactical sensitivity functions and proliferation and re-establishment terms. These equations describe the interactions between tumour cells, the uPA protease concentration and the ECM density. The systems under consideration admit exact solutions in terms of travelling wave variables under certain conditions on the model parameters. Probably, from the point of view of the biological analysis of such a complex process as the evolution and spread of a tumor, the restrictions (6)-(7), and also (16)-(18) may seem rather artificial. However, it seems interesting to us that at least under certain conditions it is possible to obtain exact solutions of such a complex systems. 

We present the exact solutions that are acceptable from a biological point of view, both for the model without proliferation and for the model with proliferation and re-establishment terms. Despite the changes made to the model of \cite{Chaplain&Lolas2006}, the profiles of these functions are very close to the profiles obtained in \cite{Chaplain&Lolas2006} from numerical computations. Due to the fact that the obtained solutions are expressed through known functions and have a quite simple form, they are easy to analyze.

We will also briefly show other exact solutions obtained formally and not suitable for biological analysis.

\end{document}